\documentclass[usenatbib, usegraphicx]{mn2e}
\usepackage{amssymb}
\usepackage{epsfig}
\usepackage{multirow}
\usepackage{tabularx}
\usepackage{color}
\usepackage{url}
\usepackage{nicefrac}

\newcommand{\ra}[3]{\mbox{#1$^h$#2$^m$#3$^s$}}
\newcommand{\dec}[3]{\mbox{#1\degr#2'#3''}}
\newcommand{\ha}{\mbox{H\,$\alpha$}}
\newcommand{\hb}{\mbox{H\,$\beta$}}
\newcommand{\fline}[2]{\mbox{$[$#1\,{\sc #2}$]$}}
\newcommand{\sfr}[1]{\mbox{#1 M$_\odot$ yr$^{-1}$}}
\newcommand{\lya}{\mbox{Ly $\alpha$}}

\author[J. Swinbank et al]{John Swinbank$^1$\thanks{j.swinbank@uva.nl}, Joanne Baker$^{2}$, Jordi Barr$^3$, Isobel Hook$^3$,
\newauthor
and Joss Bland-Hawthorn$^4$\\
$^1$Astronomical Institute ``Anton Pannekoek'', University of Amsterdam, Postbus 94249, 1090 GE Amsterdam, The Netherlands\\
$^2$Nature, 4 Crinan Street, London N1 9XW, UK\\
$^3$Astrophysics, University of Oxford, Denys Wilkinson Building, Keble Road, Oxford OX1 3RH, UK\\
$^4$Sydney Institute for Astronomy, School of Physics A28, University of Sydney, NSW 2006, Australia}
\title[Tunable filter imaging of high-$z$ quasar fields]{Tunable filter imaging of high redshift quasar fields}
\begin{document}
\maketitle
\begin{abstract}

We have used the Taurus Tunable Filter to search for Lyman $\alpha$ emitters
in the fields of three high redshift quasar fields: two at $z\sim2.2$ (MRC
B1256-243 and MRC B2158-206) and one at $z\sim4.5$ (BR B0019-1522). Our
observations had a field of view of around 35 square arcminutes, and reached
AB magnitudes of magnitudes of $\sim$21 (MRC B1256-243), $\sim$22 (MRC
B2158-206), and $\sim$22.6 (BR B0019-1522), dependent on wavelength. We have
identified candidate emission line galaxies in all three of the fields, with
the higher redshift field being by far the richest. By combining our
observations with simulations of the instrumental response, we estimate the
total density of emission line galaxies in each field. Seventeen candidate
emission line galaxies were found in within 1.5 Mpc of BR0019-1522, a number
density of $4.9 \pm 1.2 \times 10^{-3}$ Mpc$^{-3}$, suggesting a significant
galaxy overdensity at $z\sim4.5$.

\end{abstract}

\section{Introduction}
\label{sec:intro}

\begin{table*}
\begin{center}
\caption{Details of the observations made of each target. Standard stars
(HD49798, EG 274 and EG 21) for photometric calibration were observed with the
same instrumental configuration.}\label{tab:obs}
\begin{tabular}{cccccccl}
\hline
Target & Redshift & \multicolumn{2}{c}{Position (J2000)} & Date & Exposure & Axial            & Comment \\
       &          & RA & Dec                             &      & Time (s) & Wavelength (\AA) &\\
\hline
MRC B1256-243 & 2.263 & \ra{12}{59}{12.6} & \dec{-24}{36}{05} & 2003 July 27 & 15 $\times$ 60  & 3957.2 & Repeated \\
& & & & & & 3967.1 & twice. \\
& & & & & & 3977.1 & \\
& & & & & & 3987.1 & \\
\\
MRC B2158-206 & 2.249 & \ra{22}{01}{27.0} & \dec{-20}{25}{36} & 2003 July 27 & 15 $\times$ 60  & 3959.1 & Repeated \\
& & & & & & 3969.1 & four \\
& & & & & & 3979.1 & times. \\
& & & & & & 3989.0 & \\
& & & & & & 3999.0 & \\
\\
BR B0019-1522 & 4.528 & \ra{00}{22}{08.0} & \dec{-15}{05}{39} & 1997 Nov. 6 & 600 & 6709.5 & Repeated \\
& & & & & & 6725.9 & eight \\
& & & & & & 6742.3 & times. \\
\hline
\end{tabular}
\end{center}
\end{table*}

The evolution of clustering with cosmic time is widely recognised as one of
the most rigid tests of the cold dark matter paradigm \citep{Kaiser91,
Springel05}. However, locating high redshift clusters is challenging. The
traditional methods of X-ray and blind optical searches are limited: X-ray
surveys can detect only the most luminous sources at high-$z$, while optical
searches are highly vulnerable to projection effects. In order to overcome
these limitations, a way of targeting the search is needed.

Since the earliest studies, it has been established that quasars are
associated with groups and clusters of galaxies \citep{Bahcall69, Oemler72}.
More recently, \citet{McLure01} argued that a close match between the space
density of clusters and that of quasars indicates that practically all
clusters contained an AGN at high redshift. Further, \citet{Rawlings04}
propose that radio jets from AGN are a major influence on cluster evolution.
They suggest that a galaxy merger within the cluster triggers a radio-jet
episode; the jets then delivery energy to the intracluster medium, heating it
and preventing it from falling into the other developing cluster galaxies.
These galaxies are thus starved of fuel, and star formation within the cluster
will effectively shut down. \citeauthor{Rawlings04} speculate that every
protocluster undergoes such an episode, strengthening the link postulated by
\citeauthor{McLure01}.

This relationship between galaxy overdensities and AGN suggests a method for
locating high-$z$ clusters: we can use quasars as convenient `anchors' for our
search.  This technique has already been exploited by others with notable
success: for example, \citet{Stiavelli05} tentatively report the detection of
clustering around a radio-quiet quasar at $z = 6.28$.

To date most galaxy clusters detected around AGN have been identified based on
statistical overdensities of objects observed in their vicinity. A better
strategy for overcoming foreground contamination is to identify individual
star forming galaxies in the AGN field by their characteristic redshift
dependent features. In particular, Lyman $\alpha$ emission has been used to
identify high redshift galaxies for some time. Among the first high redshift
objects identified by emission lines were the $z = 4.55$ Ly $\alpha$ emitters
observed in the field of the quasar BR B2237-0607 by \citet{Hu96}. Since then,
a series of highly profitable observations of Ly $\alpha$ emitters in AGN
fields have been carried out. \citet{Kurk00} and \citet{Pentericci00} used a
combination of narrow- and broad-band imaging with follow-up spectroscopy to
identify a galaxy overdensity within 1.5 Mpc of the $z = 2.156$ radio galaxy
PKS B1138-262. Similar results have been achieved for the radio galaxies TN
J1338-1942 \citep[$z=4.1$;][]{Venemans02}, TN J0924-2201
\citep[$z=5.2$;][]{Venemans04, Overzier06} and MRC B0316-257
\citep[$z=3.13$;][]{Venemans05} and 6C0140+326 \citep[$z=4.413$;][]{Kuiper11}.

While this combination of broad and narrowband imaging has produced
demonstrably successful results, the more direct antecedents of this work have
adopted an alternative approach. The \textit{Taurus Tunable Filter} (TTF)
instrument, installed on the Anglo-Australian Telescope, provided a powerful
method of narrow-band (of order 10 \AA) imaging over a large range of
wavelengths \citep{BH982}. \citet{Bremer99} introduced the strategy used to
search for line emitters at a given redshift with TTF: broadly, the tunable
filter is stepped across a range of wavelengths around the expected redshifted
position of the emission. Emission line galaxies then appear brighter in those
frames centred on the spectral line.

Considerable success has been achieved at lower redshifts with this technique.
\citet{Baker01} located a cluster around the $z = 0.9$ radio-loud quasar MRC
B0450-221 using TTF to search for $[$O\,{\sc ii}$]$ 3727 \AA{} emission. The
same technique was used by \citet{Barr04}, who examined six radio-loud quasars
at redshifts $0.8 < z < 1.3$, identifying a total of 47 candidate emission
line galaxies (ELGs), at an average space density around 100 times higher than
that found locally.

Further work with TTF was performed by \citet{Francis04}, who targeted Ly
$\alpha$ emitters within 1 Mpc of the $z=2.159$ radio loud quasar PKS
B0424-131 without making {\it any} detections. These authors selected this
extremely luminous UV source with the expectation of finding Ly $\alpha$
fluorescent clouds in the vicinity of the quasar but these were not detected.
With specific application to PKS B0424-131, \citet{Bruns11} demonstrated that
the most intrinsically UV-luminous quasars observed beyond $z=1$ suppress star
formation in low-mass haloes ($M_{\rm vir} \lesssim 10^{12}$ M$_\odot$) within
a megaparsec of the quasar. The intense UV radiation field is expected to
photo-evaporate HI clouds which presumably accounts for the lack of
detections.  We return to this point in our conclusion
(\S~\ref{sec:conclusion}).

The present work continues to push TTF to higher redshifts, searching three
quasar fields at redshifts up to $z \sim 4.5$. The objects selected include
examples of both radio-loud and radio-quiet quasars, and their environments
are compared. Section \ref{sec:obs} of this paper describes the observations,
including target selection, instrumental characteristics and a note on data
reduction. Section \ref{sec:sim} describes simulations performed to examine
statistical properties and completeness of our sample. Section \ref{sec:id}
describes how candidate ELGs were identified and presents details on the
detections, as well as considering the possible sources of mis-identified
`interloper' objects. Section \ref{sec:properties} analyses the distribution
and properties of the sample. Our conclusions are summarised in Section
\ref{sec:conclusion}.  Throughout, we assume an $H_0 = 70$ km s$^{-1}$
Mpc$^{-3}$, $\Omega_{\Lambda} = 0.7$, $\Omega_{\mathrm{M}} = 0.3$ cosmology.

\section{Observations}
\label{sec:obs}

\subsection{Target selection}

Two data sources were used for this analysis. The authors used TTF to observe
objects drawn from the Molonglo Quasar Sample \citep[MQS;][]{Kapahi98} of
low-frequency-selected radio-loud quasars in July 2003. Six targets had been
selected from the MQS on the basis of observability, suitable redshifts being
limited by the necessity to place Lyman $\alpha$ within the wavelength ranges
accessible to TTF's order-blocking filters. Due to weather constraints, only
two quasars were observed: MRC B1256-243 ($z = 2.263$) and MRC B2158-206 ($z =
2.249$). Immediately following each quasar observation, a standard star was
observed with the same instrumental settings for flux calibration.  In
addition, observations of BR B0019-1522, a $z = 4.528$ radio-quiet quasar,
were drawn from the Anglo-Australian Observatory archive. These data were
taken on 1997 November 6 by Bland-Hawthorn, Boyle and Glazebrook, and were
accompanied by companion observations of a standard star.  Details of each
target are given in Table \ref{tab:obs}.

\subsection{Instrumental setup and characteristics}

Throughout this work, a distinction is drawn between a \textit{frame}
(corresponding to one set of data read from the CCD), an \textit{image} (a
number of frames at the same etalon settings which have been combined for
analysis) and a \textit{field}, or stack of images of the same area of sky at
different etalon settings.

\subsubsection{Wavelength variation and the optical axis}
\label{sec:wlvariation}

Fabry-P\'erot images have a quadratic radial wavelength dependence of the form
$\lambda_\theta = \lambda_{centre}(1 - \theta^2/2)$ \citep{Bland89}, where
$\theta$ is the off-axis angle at the etalon. In a typical observation, the
wavelength varies across the field by around 1\% of $\lambda_{centre}$.
Wavelength calibration is performed with respect to the axial wavelength; for
any given pixel position on the image, it is then possible to calculate the
wavelength observed at that point.

\subsubsection{Objects at $z \sim 2.2$}

The TTF was used at $f/8$ on the AAT in combination with the EEV2 CCD. This
resulted in a scale of 0.33'' per pixel. After processing, the total useful
rectangular field of view in the observations was around 7' by 5'. The radial
wavelength variation described in Section \ref{sec:wlvariation} resulted in a
shift of 1.4~\AA{} at 2' from the optical axis and 6.7~\AA{} at 4' from the axis.
Conditions were photometric, and seeing was on the order of 1.5''. The full
width at half maximum of the etalon transmission band was 7.5~\AA.

\begin{figure}
\begin{center}
\includegraphics{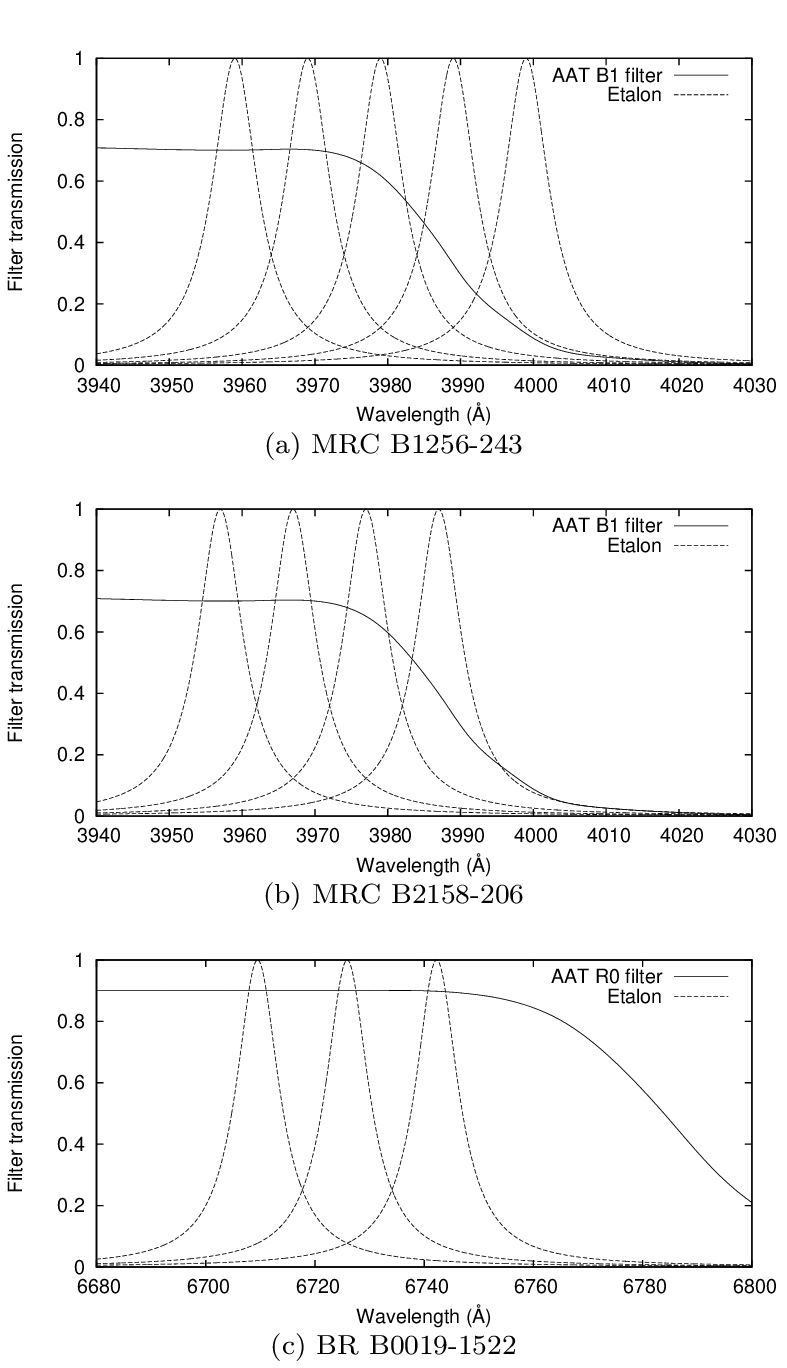}
\caption{On-axis etalon transmission bands for each of the three fields
observed shown relative to the relevant order-blocking filter used on the
telescope. Away from the optical axis the etalon transmission shifts to
shorter wavelengths (\S\ref{sec:wlvariation}).}\label{fig:trans}
\end{center}
\end{figure}

The targets were scanned at etalon plate spacings corresponding to a series of
wavelength steps of approximately 10~\AA, the aim being to straddle the
redshifted Ly $\alpha$. However, an intermediate-band order-blocking filter is
necessary to eliminate unwanted wavelengths and other orders of interference.
In this case, the AAT's B1 filter was the best available.  Unfortunately, the
observed wavelengths were at the very edge of the filter transmission, as
shown in Fig. \ref{fig:trans}: the signal to noise ratio therefore decreases
significantly with wavelength. Table \ref{tab:obs} and Fig. \ref{fig:trans}
record observations of MRC B1256-243 at 3987.1 \AA.  When these data were
analysed, it was clear that the reduced filter transmission had resulted in no
useful results at this wavelength. These data are not considered further in
this work. The MRC B2158-206 observations at 3989.0 \AA{} and 3999.0 \AA{} are
included hereafter, but did not include any useful detections.

Each CCD frame contained a total of 30 minutes of observations, taken at two
separate axial wavelengths. Each wavelength was exposed for 60 seconds a total
of 15 times. This procedure was repeated twice in the case of MRC B1256-243
and four times for MRC B2158-206; the total exposure times at each wavelength
are thus 30 minutes and 1 hour, respectively. Between each image, the
telescope pointing was shifted slightly: this enabled the easy identification
and subsequent elimination of diametric ghosts in the data.

\subsubsection{Objects at $z \sim 4.5$}

The TTF was used at $f/8$ on the AAT in combination with the MITLL2 CCD. This
resulted in a scale of 0.37'' per pixel. After processing, the total useful
rectangular field of view in the observations was 9'17'' by 4'10''.  The
radial wavelength variation described in Section \ref{sec:wlvariation}
resulted in a shift of 5.1~\AA{} at 2' from the optical axis and 20.3~\AA{} at
4' from the axis.  Conditions were photometric, and the seeing was on the
order of 1.5". The full width at half maximum of the etalon transmission band
was 9.5~\AA. The AAT's R0 intermediate-band order-blocking filter was used:
this provided effectively constant transmission across the wavelength range
under consideration.

Each CCD frame contained a total of 30 minutes of observations: ten at each of
three axial wavelengths. Eight CCD frames were recorded, resulting in a total
of 80 minutes exposure for each axial wavelength. As before, the telescope
position was shifted slightly between images.

\subsection{Data reduction and catalogue construction}

Data reduction proceeds broadly as for standard broadband imaging. A full
consideration of the issues surrounding tunable filter data is given by
\citet{Jones012} and \citet{Jones02}. The various different images of each
field at the same axial wavelengths were aligned by a marginal centroid fit on
bright stars and then combined. Wavelength calibration was performed through
an emission line, as described by \citeauthor{Jones02}; xenon and
copper-helium arc lamps were used for the $z \sim 2.2$ fields, and a neon arc
lamp for BR B0019-1522.

After the data had been reduced, object detection and fixed aperture
photometry were performed on each image using {\sc SExtractor}
\citep{Bertin96}. The object detection parameters were defined as described in
the next section.

\subsection{Photometry}
\label{sec:photo}

The observations of the standard stars were reduced in the same way. For each
star, {\sc SExtractor} was used to perform aperture photometry yielding a
count $C_\mathrm{s}$. This corresponds to a known magnitude $m_\mathrm{s}$,
based on \citet{Hamuy92} for the lower redshift fields or from the ESO
Standard Star Catalogue for that of BR B0019-1522. If the exposure time on the
standard is $t_\mathrm{s}$ and that on an object in the field is
$t_\mathrm{Obj}$, the AB magnitude of the object is

\begin{equation}
m_\mathrm{AB} = m_\mathrm{s} - 2.5 \log_{10} (C_\mathrm{Obj}t_\mathrm{s})/(C_\mathrm{s}t_\mathrm{Obj}).
\end{equation}

The AB magnitude system \citep{Oke74} is defined by $m_\mathrm{AB} = -2.5
\log_{10} f_\nu - 48.60$ where $f_\nu$ is the flux in units of \mbox{ergs
cm$^{-2}$ s$^{-1}$ Hz$^{-1}$}.  The monochromatic flux $f_\lambda$, in units of
\mbox{ergs cm$^{-2}$ s$^{-1}$ \AA$^{-1}$}, is then

\begin{equation}
\label{eq:abtoflux}
f_\lambda = (c \times 10^{-\left(m_{\mathrm{AB}} + 48.60\right)/2.5})/\lambda^2.
\end{equation}

Conversion from $f_\lambda$ to the total flux in the band, $f_\mathrm{total}$
is performed by multiplying by the effective width of the etalon transmission.
The etalon transmission band may be taken as Lorentzian, normalised to 1 at
the wavelength of peak transmission, thus:

\begin{equation}
\label{eq:ttfpass}
T(\lambda) = (\lambda_{\nicefrac{1}{2}}^2 / 4)/((\lambda - \lambda_\mathrm{c})^2 + \lambda_{\nicefrac{1}{2}}^2 / 4)
\end{equation}

where $\lambda$ is the wavelength, $\lambda_c$ the central wavelength of the
band and $\lambda_{\nicefrac{1}{2}}$ its full width at half maximum. Assuming
that $\lambda_\mathrm{c} \gg \lambda_{\nicefrac{1}{2}}$, Equation
\ref{eq:ttfpass} may be integrated over $0 \le \lambda \le \infty$ to yield a
width of $\pi \lambda_{\nicefrac{1}{2}}/2$. Combining this with Equation
\ref{eq:abtoflux} yields a total flux in the band of

\begin{equation}
\label{eq:fluxinband}
f_{\mathrm{total}} = (\pi c \lambda_{\nicefrac{1}{2}} \times 10^{-\left(m_\mathrm{AB} + 48.60\right)/2.5})/2 \lambda_\mathrm{c}^2
\end{equation}

with units \mbox{ergs cm$^{-2}$ s$^{-1}$}.

It is worth noting that this measures the flux received in the etalon
passband, and is thus a lower limit of the line flux of the ELG: variations of
line shapes and widths, and their positions relative to the etalon passband,
will cause the fluxes measured to be systematically underestimated. They
should therefore be regarded as lower limits.

\section{Simulations}
\label{sec:sim}

\begin{figure*}
\begin{center}
\includegraphics{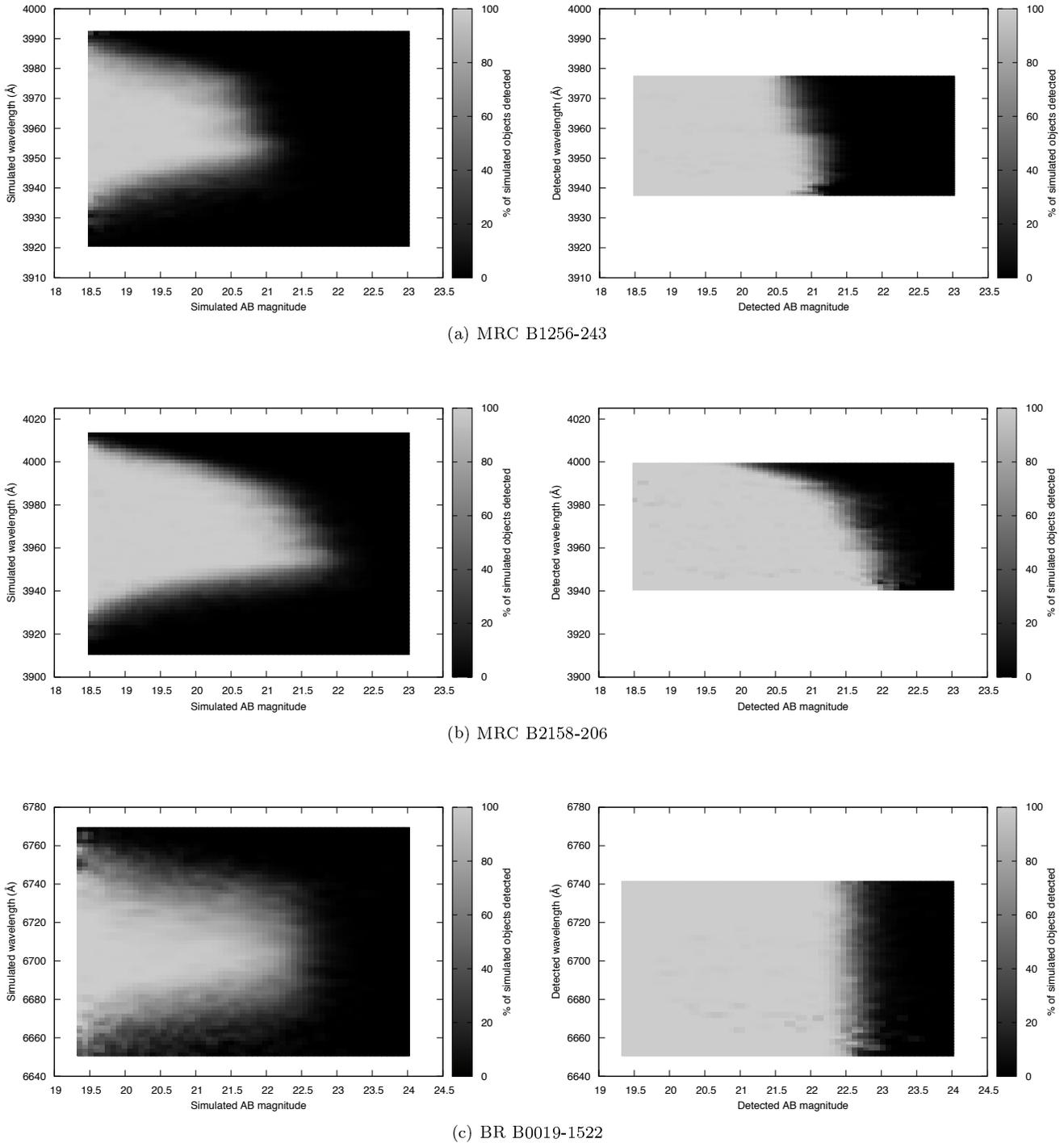}
\caption{Depths of each of the three fields as determined by the simulations
described in Section \ref{sec:dof}. On the left, the data is plotted in terms
of simulation inputs; on the right, in terms of the measurements made from
the simulated images. Note that the effects of the blocking filter are clearly
seen in the two upper (lower redshift) fields, as the completeness tails off
at higher wavelength. The higher redshift BR B0019-1522 field falls well
within the blocking filter, so the depth is relatively constant with
wavelength across the observed range.}
\label{fig:simresults}
\end{center}
\end{figure*}

We constructed a series of simulated images: data with properties similar to
our observations, but containing a known population of objects. The analysis
of these enables us to address the following questions:

\begin{itemize}
\item What are the most appropriate {\sc SExtractor} parameters for
extracting useful data from the images?
\item To what depth is each field complete--and how does that vary over the
field?
\item To what extent is our analysis prone to mis-identifying spurious `noisy'
features in an image as candidate emission line galaxies?
\end{itemize}

\subsection{Construction of simulated images}

Images were simulated in two stages: first, a background was generated, then
objects were superimposed on top of it.

Due to the properties of the blocking filter and the variation of wavelength
across the image, the background signal is not constant across the image. Each
data image was therefore divided into 100 by 100 pixel blocks, and the mean
background signal and associated noise was measured in each block. Simulated
blocks were then generated matching each of these, and then recombined to form
an overall simulated background of the same shape as the data.

A Ruby\footnote{\url{http://www.ruby-lang.org/}} code was written to simulate
the expected properties of objects we might observe. Objects were simulated at
random redshifts (over the range the observations might be expected to cover)
and pixel positions within the images. Based on the work of
\citet{LeDelliou06}, our observations were not expected to be sensitive to
continuum emission from ELGs, so this was not considered. Further, the ELGs
are spatially unresolved, so were simulated with a Gaussian point spread
function equal to the measured seeing. An emission line model was developed
based on the widths and profiles of high-$z$ Lyman $\alpha$ emitters based
chiefly on the $z \sim 4.5$ objects observed by \citet{Dawson04}.
Experimentation suggested that the results obtained were not sensitive to line
profile; velocity widths in the range 100--1000 km\,s$^{-1}$ were chosen
based on both \citet{Dawson04} and the more extreme example documented by
\citet{Tapken04}.

The effects of the instrument on the objects' detectabilty were then
considered before they were added to the background images. First a correction
for the order-blocking filter transmission was applied, using the position of
the object within the field to determine the observed wavelength and hence
filter transmission.  The line profile was then multiplied by the transmission
profile of the etalon for the image under construction.

\subsection{Results of simulations}

Following the procedure above, simulations were run of all three fields. For
each data image, a total of 500 simulated images were constructed, each
containing 500 simulated sources.

\subsubsection{Detection parameters}
\label{sec:detpar}

Source extraction was run multiple times on each image with different
{\sc SExtractor} configuration parameters. In each case, the results were
compared with the catalogue of simulated objects in the image. The combination
of parameters that produced the greatest number of detections of known objects
combined with the smallest number of spurious detections of noise were then
used for the analysis of both the simulations and the observed data. These
parameters are listed in Table \ref{tab:sextractor}.

\begin{table}
\begin{center}
\caption{Optimal {\sc SExtractor} parameters determined by simulations and
used throughout this work.}\label{tab:sextractor}
\begin{tabular}{ccp{4.1cm}}
\hline
Parameter & Value & Description \\
\hline
{\sc detect\_minarea} & \phantom{0}6\phantom{.0} & Minimum number of pixels per detection. \\
{\sc detect\_thresh} & \phantom{0}1.3 & Detection threshold in $\sigma$ above local background. \\
{\sc back\_size} & 64\phantom{.0} & Size in pixels of mesh used for background estimation. \\
{\sc phot\_apertures} & \phantom{0}6\phantom{.0} & Aperture diameter (pixels). \\
\hline
\end{tabular}
\end{center}
\end{table}

\subsubsection{Depths of fields}
\label{sec:dof}

As in the previous section, a source detection procedure was run on each
image and the results compared with the known simulation inputs. This time,
the fraction of the objects at each wavelength and magnitude which were
detected was recorded. The results are shown Fig. \ref{fig:simresults}.

Note that this data can be recorded both in terms of the \textit{simulated}
wavelength and magnitude and and their \textit{detected} equivalents. For any
given pixel position in a field, an object can only be detected as peaking at
one of a limited range of wavelengths, since its peak will be seen to appear
at the wavelength of the image in which it occurs (of which there are at most
5). Hence, an object which is simulated with a very bright magnitude, but at a
wavelength far from the peak transmission of any of the filters, will be
detected with a somewhat dimmer magnitude at a wavelength corresponding to the
image in which it is brightest. Fig. \ref{fig:simresults} shows both the
simulated (on the left) and detected (on the right) quantities for each of
the three fields.

\section{Identification of candidate ELGs}
\label{sec:id}

\begin{table*}
\begin{center}
\caption{ELG candidates in the field of BR B0019-1522. The AB magnitude given
is that measured in the peak from with no correction for galactic extinction
or etalon transmission; the flux is calculated from that magnitude via Equation
\ref{eq:fluxinband}.}\label{tab:elgresults}
\begin{tabular}{lccccccc}
\hline
Field & ELG & \multicolumn{2}{c}{Position (J2000)} & Projected distance & Lyman $\alpha$ Peak & AB   & Flux in band \\
      & Id. &  R.A. & Decl.                        & from Quasar (Mpc)  & Wavelength (\AA)    & mag. & (ergs cm$^{-2}$ s$^{-1} \times 10^{18}$)\\
\hline
MRC B1256 & A & \ra{12}{59}{23.2} & \dec{-24}{37}{32.9} & 1.428 & 3966 & 20.9 & 371 \\
          & B & \ra{12}{59}{15.7} & \dec{-24}{37}{40.7} & 0.871 & 3966 & 21.1 & 293 \\
          & C & \ra{12}{59}{02.7} & \dec{-24}{37}{15.1} & 1.257 & 3957 & 20.9 & 363 \\
          & D & \ra{12}{59}{05.3} & \dec{-24}{37}{31.3} & 1.085 & 3960 & 20.7 & 424 \\
\\
MRC B2158 & A & \ra{22}{01}{26.0} & \dec{-20}{25}{08.0} & 0.263 & 3956 & 21.8 & 161 \\
          & B & \ra{22}{01}{41.7} & \dec{-20}{24}{03.5} & 1.986 & 3971 & 21.7 & 192 \\
\\
BR B0019 & A & \ra{0}{21}{56.9} & \dec{-15}{04}{04.3} & 1.229 & 6673 & 22.5 & \phantom{0}37 \\
         & B & \ra{0}{22}{03.8} & \dec{-15}{07}{41.2} & 0.898 & 6706 & 22.5 & \phantom{0}37 \\
         & C & \ra{0}{22}{08.8} & \dec{-15}{06}{58.8} & 0.531 & 6705 & 22.0 & \phantom{0}57 \\
         & D & \ra{0}{22}{08.8} & \dec{-15}{06}{56.3} & 0.515 & 6704 & 21.7 & \phantom{0}71 \\
         & E & \ra{0}{21}{57.8} & \dec{-15}{06}{58.7} & 1.105 & 6697 & 22.7 & \phantom{0}31 \\
         & F & \ra{0}{22}{14.5} & \dec{-15}{06}{42.6} & 0.748 & 6717 & 22.1 & \phantom{0}52 \\
         & G & \ra{0}{22}{12.4} & \dec{-15}{06}{17.8} & 0.491 & 6716 & 22.1 & \phantom{0}51 \\
         & H & \ra{0}{22}{12.7} & \dec{-15}{06}{01.4} & 0.471 & 6697 & 22.5 & \phantom{0}37 \\
         & I & \ra{0}{22}{07.6} & \dec{-15}{05}{27.1} & 0.087 & 6694 & 22.4 & \phantom{0}39 \\
         & J & \ra{0}{21}{58.6} & \dec{-15}{04}{56.2} & 0.940 & 6701 & 22.3 & \phantom{0}43 \\
         & K & \ra{0}{22}{14.2} & \dec{-15}{04}{20.6} & 0.785 & 6680 & 22.6 & \phantom{0}32 \\
         & L & \ra{0}{22}{14.8} & \dec{-15}{07}{22.1} & 0.939 & 6719 & 22.5 & \phantom{0}37 \\
         & M & \ra{0}{22}{15.3} & \dec{-15}{06}{52.7} & 0.849 & 6716 & 22.2 & \phantom{0}48 \\
         & N & \ra{0}{22}{11.5} & \dec{-15}{05}{04.1} & 0.405 & 6706 & 22.3 & \phantom{0}43 \\
         & O & \ra{0}{22}{18.0} & \dec{-15}{04}{36.8} & 1.038 & 6694 & 22.4 & \phantom{0}39 \\
         & P & \ra{0}{21}{53.9} & \dec{-15}{05}{58.2} & 1.351 & 6685 & 22.4 & \phantom{0}40 \\
         & Q & \ra{0}{22}{13.9} & \dec{-15}{05}{08.8} & 0.597 & 6689 & 22.5 & \phantom{0}35 \\
\hline
\end{tabular}
\end{center}
\end{table*}

\begin{figure}
\begin{center}
\includegraphics{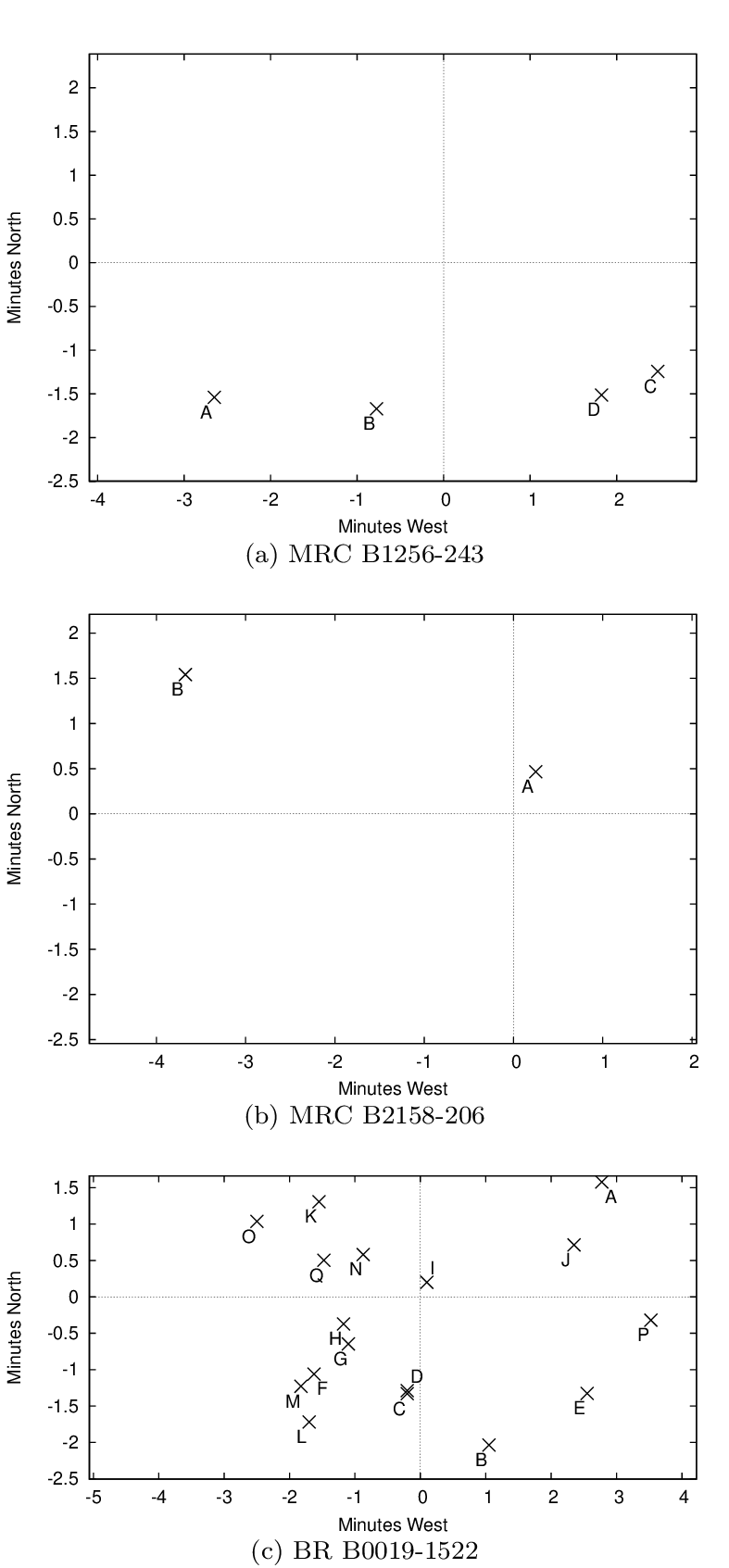}
\caption{Relative positions of the ELG candidates detected in each of the
three fields. The dimensions of the plots indicate the size of the observed
fields. The quasars are located at the origin. The letters refer to the ELG
designations used throughout the text.}\label{fig:elgcandidates}
\end{center}
\end{figure}

{\sc SExtractor} was used with the parameters determined in Section
\ref{sec:detpar} and a detection threshold of 5$\sigma$ to build a catalogue
of sources for each image. Within each field, the catalogues from each image
were cross-matched: objects were associated by position, with a three pixel
threshold.

These observations are not deep enough to observe continuum flux from a
typical Lyman $\alpha$ emitting galaxy \citep{LeDelliou06}. Given the likely
range of line widths \citep{Dawson04, Tapken04}, we do not expect to observe
Lyman $\alpha$ emitters in more than two adjacent passbands. Objects which
were identified in either one or two bands were therefore flagged for further
investigation.

In order to minimise the risk of contamination by noisy artefacts, all
flagged objects were examined eye, and those which appeared unphysical or
corresponded to sites of corruption by (for example) heavy cosmic ray
or charge trapping activity in the original images were rejected.

\subsection{MRC B1256-243}

Four candidate emission line galaxies were identified in the field of MRC
B1256-243. Details are given in Table \ref{tab:elgresults}, and their
locations are shown in Fig. 3(a). Thumbnail images of the
candidate galaxies from each field, together with the measured fluxes, are
shown in Fig. \ref{fig:1256objects}.

\subsection{MRC B2158-206}

Two candidate emission line galaxies were identified in the field of MRC
B2158-206. Details are given in Table \ref{tab:elgresults}, and their
locations are shown in Fig. 3(b). Thumbnail images of the
candidate galaxies from each field, together with the measured fluxes, are
shown in Fig. \ref{fig:2158objects}.

\subsection{BR B0019-1522}

Seventeen candidate emission line galaxies were identified in the field of BR
B0019-1522. Details are given in Table \ref{tab:elgresults}, and their
locations are shown in Fig. 3(c). Thumbnail images of the
candidate galaxies from each field, together with the measured fluxes, are
shown in Fig. \ref{fig:0019objects}.

\subsection{Contaminants}

This section briefly addresses the likelihood that our method might
incorrectly identify another sort of object as an ELG.

\subsubsection{Continuum objects}

As per Figs. \ref{fig:trans} and \ref{fig:simresults}, the sensitivity of
our instrument varies from image to image. Therefore, it is possible that a
flat-spectrum continuum object may be detected in some images but not others,
thereby appearing to be a potential ELG.

We use the results of Section \ref{sec:sim} to estimate the probability of
this occurring. Each of the 250,000 simulated objects was sorted into one of
3,600 bins by wavelength and magnitude (each bin covering 1 \AA{} and 0.1
magnitudes). It is then possible to calculate the completeness of the bin
(i.e. the fraction of simulated objects which were recovered). Each candidate
ELG is assigned to a bin, and we then check the corresponding bins in adjacent
images for completeness. A low completeness value in these bins indicates that
a flat-spectrum object may have been `lost'.

This procedure calls into question four objects: A in the field of MRC
B2158-206, B in the field of MRC B2156-243 and E and K in the field of BR
B0019-1522. These sources were examined by eye, but there is no indication of
a faint detection in the crucial frame. They have not, therefore, been
excluded from this analysis.

\subsubsection{Lower redshift interlopers}

Another possibility is other emission lines at lower redshift may appear in
our observations. The lines which might be observed are listed in Table
\ref{tab:interlopers}.

\begin{table*}
\begin{center}
\caption{Potential low-redshift `interloper' emission lines, together with the
redshifts at which they appear and the estimated number observed in each of
the fields. The flux of each line relative to \ha{} in
a ``typical'' galaxy is given, based on \citet{Kennicutt92}.}\label{tab:interlopers}
\begin{tabular}{ccccccccccc}
\hline
Line & \AA           & Flux  & \multicolumn{2}{c}{MRC B2158-206} & \multicolumn{2}{c}{MRC B1256-243} & \multicolumn{2}{c}{BR B0019-1522} \\
     & (rest)        & ratio & $z$ & Number                      & $z$ & Number                      & $z$ & Number \\
\hline
\fline{O}{ii} & 3727 & $0.41\pm0.21$  & 0.065 & \phantom{$^*$}0.05\phantom{$^*$} & 0.060 & 0.02 & 0.803 & 1.93 \\
\hb & 4860 & $0.14\pm0.06$            & -     & -        & -     & -    & 0.383 & 1.68 \\
\fline{O}{iii} & 5007 & $0.20\pm0.15$ & -     & -        & -     & -    & 0.342 & 1.41 \\
\ha & 6548 & $1.00\pm0.00$            & -     & -        & -     & -    & 0.027 & \phantom{$^*$}0.01\phantom{$^*$} \\
\fline{N}{ii} & 6583 & $0.43\pm0.16$  & -     & -        & -     & -    & 0.021 & \phantom{$^*$}0.01\phantom{$^*$} \\
\hline
\end{tabular}
\end{center}
\end{table*}

\begin{figure}
\begin{center}
\rotatebox{270}{\resizebox{!}{\columnwidth}{\includegraphics{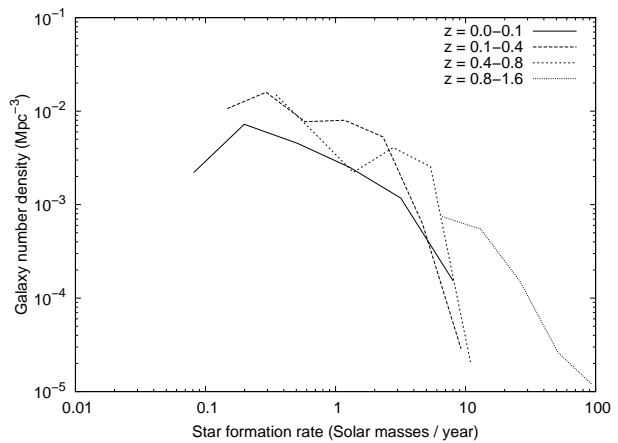}}}
\caption{Variation of galaxy number density with star formation rate for a
range of redshifts. Based on data from \citet{Cowie97} and \citet{Gallego95}.}
\label{fig:sfgs}
\end{center}
\end{figure}

\citet{Cowie97} and \citet{Gallego95} provide number density counts for star
forming galaxies at a range of redshifts. Both adopt a \mbox{$H_0 =
50$ km\,s$^{-1}$\,Mpc$^{-3}$}, $\Omega_{\Lambda} = 0$, $\Omega_{\mathrm{M}} =
1$ cosmology, which we converted to match that used in this work (Section
\ref{sec:intro}). In addition, \citeauthor{Gallego95} assume a \citet{Scalo86}
IMF; \citeauthor{Cowie97} provide a conversion to a \citet{Salpeter55} IMF,
and it is these results we adopt in this work. Based on these, we can estimate
the number density of star forming galaxies along our line of sight: see
Fig. \ref{fig:sfgs}.

\citet{Kennicutt98} provides a conversion between star formation rate in a
galaxy and \ha{} luminosity; the ratios given in Table \ref{tab:interlopers}
make it possible to convert that into expected luminiosities for the other
lines. After applying a correction for instrumental effects and galactic
extinction \citep{Schlegel98}, a locus of points in the magnitude-wavelength
completeness diagrams (Fig. \ref{fig:simresults}) on which each line at a
given redshift might be detected is determined. This locus is then integrated
to estimate the total volume over which the line might be observed at this
redshift. This procedure is then repeated along the full length of the curves
shown in Fig. \ref{fig:sfgs}. In this way, the total number of interlopers
which might be observed is estimated. The results are shown in Table
\ref{tab:interlopers}.

It is clear that the estaimted number of interlopers is negligible in the case
of the two lower-redshift fields. However, it is possible that as many as five
of the candidate ELGs in the BR B0019-1522 field are, in fact, low redshift
interlopers. This could only be confirmed by further observations.

\section{Properties of candidate ELGs}
\label{sec:properties}

In this section, we consider the distribution of candidate ELGs around the
quasars to determine whether the quasar lies in an identifiable overdensity
relative to the field.

The small number of candidates around the lower-$z$ quasars renders a
meaningful statistical analysis of the individual fields unreliable. In an
attempt to mitigate this, and given the apparent similarity of the fields,
they are both considered as one unit in this section.

\begin{figure*}
\begin{center}
\includegraphics{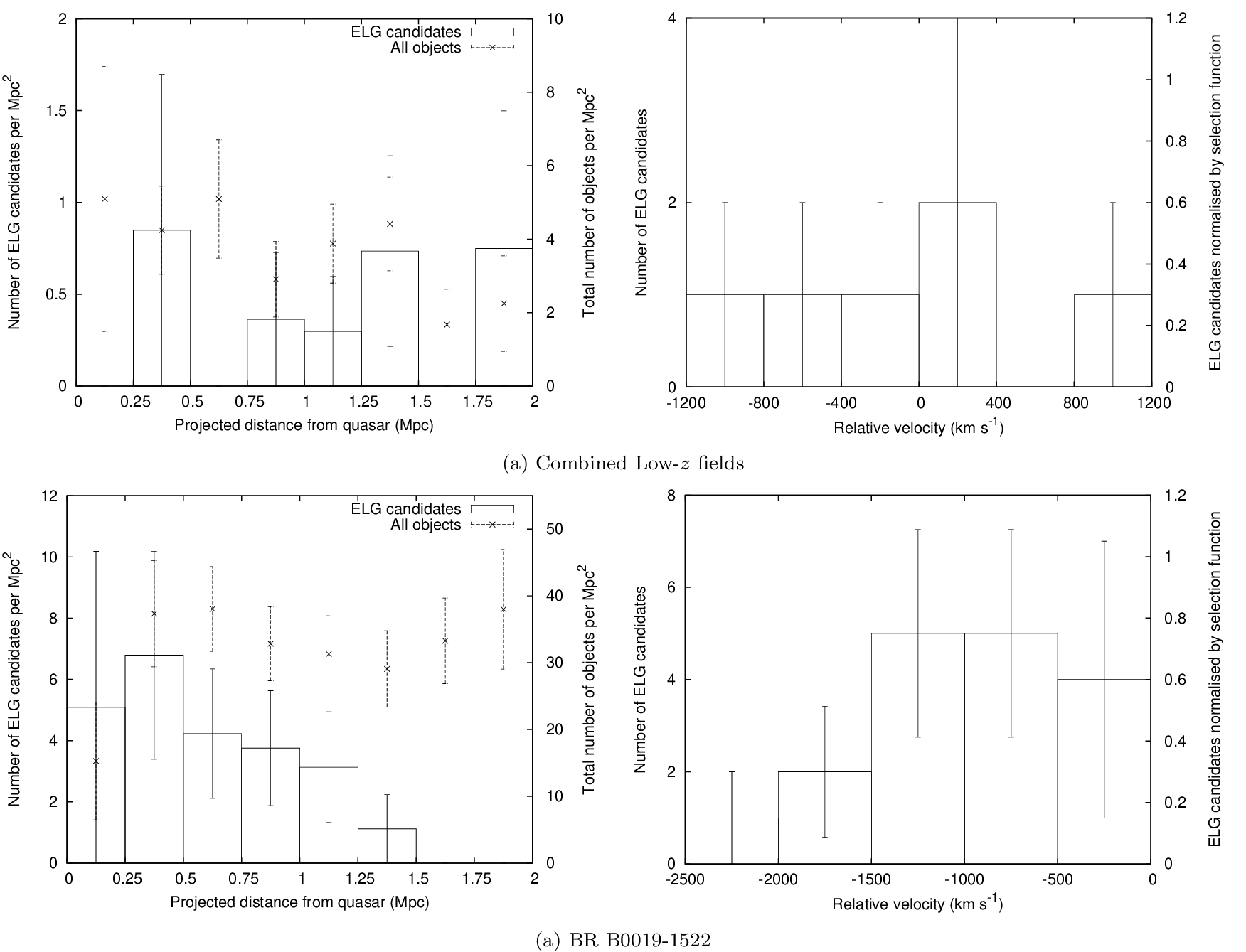}
\caption{Distribution of ELG candidates around the quasars. On the left, the
projected distance seen on the sky for both the ELG candidates (boxes) and all
the objects observed (crosses); at right, the relative velocities.}
\label{fig:distribution}
\end{center}
\end{figure*}

The distribution of ELG candidates around the quasar is shown in both
projection on the sky (left) and velocity distribution (right) in Fig.
\ref{fig:distribution}.  When calculating the projection on the sky, we have
normalised the total visible area on the sky in each distance bin. We also
plot the distribution of all objects detected by {\sc SExtractor} in the field
for comparison.

Based on these figures, there is little evidence of projected clustering in
the low-$z$ fields. However, there is a notably higher density of objects
within 1 Mpc (projected) of BR B0019-1522. This is consistent with what one
might expect from an examination of Fig. \ref{fig:elgcandidates}: note the
large number of objects to the east of the quasar in Fig. 3(c).  It is also
in-line with the scale lengths observed in clusters around other AGN
\citep{Venemans02, Bremer02, Barr04}.

There is no suggestion of clustering in velocity space in Fig.
\ref{fig:distribution}. In part, this may be due to the low number of
detections in the low-$z$ fields. In the field of BR B0019-1522, we note that
all candidates were observed as bluer than the quasar itself; this is
noteworthy, but not implausible given the wavelength range probed (6650--6740
\AA, with the quasar at 6722 \AA).  Although the bluest velocity bins show a
lower number of total counts, this can be attributed to the reduced
instrumental sensitivity at the relevant wavelengths (see Fig. 3(c)).

The space density of galaxies in the three fields may also be estimated.  As
alluded to in the previous section, the comoving volume being probed by our
measurements varies with wavelength and magnitude. Consider for example Fig.
2(a): a bright object--magnitude 19, say--may be detected at a range of
wavelengths, from around 3920 \AA{} to 4010 \AA. A fainter object at, for
instance, magnitude 22 is only detected if it lies within a much smaller
wavelength range: around 3940 \AA{} to 3960 \AA. Therefore, we define an
`accessible volume', $\mathcal{V}_n$, for each detected object $n$ within the
field. $\mathcal{V}_n$ is calculated by taking the locus of points in Fig.
\ref{fig:simresults} occupied by a source with the observed properties and
integrating over all wavelengths. The density is taken as $\rho =
1/\mathcal{V}_1 + 1/\mathcal{V}_2 + ... + 1/\mathcal{V}_n$. The results for
our fields are given in Table \ref{tab:density}.

\begin{table}
\begin{center}
\caption{Estimated space and star formation rate densities, together with the
total number of ELG candidates (\#), for each of the fields
observed. Note that our observations are valid only to an approximately
defined lower limit of star formation.}\label{tab:density}
\begin{tabular}{cccc}
\hline
Field & \# & Number density             & SFR density \\
      &   & (Mpc$^{-3}\,\times\,10^4$) & (M$_\odot\;$\,yr$^{-1}$\,Mpc$^{-3}$) \\
\hline
MRC B1256 & \phantom{0}4 & $22.48 \pm 11.64$ & $0.0346 \pm 0.0174$ \\
MRC B2158 & \phantom{0}2 & $\phantom{0}9.09 \pm \phantom{0}6.52$ & $0.0070 \pm 0.0049$ \\
BR B0019  &           17 & $49.09 \pm 12.21$ & $0.0484 \pm 0.0117$ \\
\hline
\end{tabular}
\end{center}
\end{table}

It is also instructive to estimate the star formation rates found in these
fields. Based on \citet{Kennicutt94} combined with \citet{Brocklehurst71} and
\citet{Hu96}, we arrive at the relationship:

\begin{equation}
\mathrm{SFR}(\mathrm{M}_\odot\,\mathrm{yr^{-1}}) = 0.91 \times 10^{-42} L(\mathrm{Ly} \alpha) (\mathrm{erg\,s^{-1}})
\label{eq:sfr}
\end{equation}

It should be noted that \lya{} is a very poor indicator of star formation
rate. It is resonantly scattered by neutral hydrogen, and hence has a high
chance of absorption either before leaving the galaxy or by clouds in the
intergalactic medium \citep{Haiman99}. Further, \citet{VG93} argues that \lya{}
emission in starbursts is strongly dependent on the age of the burst,
rendering the calibration of Equation \ref{eq:sfr} unreliable from around
$10^7$ years after the burst start. Nevertheless, \lya{} is the only
diagnostic available to us, so we persist in these estimates with caution.

We take the star formation rate density as $\rho_{SFR} = SFR_1/\mathcal{V}_1 +
SFR_2/\mathcal{V}_2 + ... + SFR_n/\mathcal{V}_n$, where $SFR_n$ is the star
formation rate associated with ELG candidate $n$ as calculated by Equation
\ref{eq:sfr}. Recall from Section \ref{sec:photo} that the line fluxes are
systematically underestimated since objects will fall outside the peaks of the
etalon passpands. Making the approximation that objects are evenly spread in
wavelength around the etalon peaks, we apply a correction to the observed
magnitudes of 0.23 (in the low-$z$ field) or 0.27 (BR B0019-1522 field) to
account for this.  We correct the results for completeness based on Fig.
\ref{fig:simresults}: a single detection in an area with a low detection rate
is taken as representative of a larger population.

The results are shown in Table \ref{tab:density}. Note that our observations
are sensitive to galaxies only down to some minimum level of star formation
(\sfr{9} in the case of MRC B2158-206 and BR B0019-1522; \sfr{25} in the case
of MRC B1256-243): there may be a fainter population which we do not probe.

It is noteworthy that the star formation rate in the field of MRC B1256-243 is
anomalously high, but the large uncertainties in the field and the higher
minimum detectable rate render this result questionable. The most well
constrained result is that for BR B0019-1522; our results there are broadly
similar to those reported by \citet{Venemans02} around the $z = 4.1$ radio
galaxy TN J1338-1942. In all three fields, the number of objects detected is
higher than that which might be expected in the absence of any clustering.
Based on \citet{Cowie97}, we might expect on average 0.86 galaxies in the
field of MRC B2158-206, 0.25 in that of MRC B1256-243, and 1.3 in that of BR
B0019-1522, while an extrapolation from the results of the LALA \citep[`Large
Area Lyman $\alpha$';][]{Rhoads00} survey suggests we should observe 1.1
objects in the field of MRC B2158-206, 0.8 in that of MRC B1256-243 and 2.1 in
that of BR B0019-1522 (assuming that the density of \lya{} emitters is similar
at $z \sim 2.2$ to that observed at $z \sim 4.5$).

\section{Conclusions}
\label{sec:conclusion}

Until recently, it has proved difficult to find high-redshift clusters and,
indeed, there are very few known beyond $z \sim 1$.  The detection of hot
X-ray emission from intracluster gas followed by optical imaging and/or
spectroscopic confirmation becomes inefficient for detecting more distant
clusters; a manifestly higher success rate is achieved by targeting the
vicinity of high redshift radio galaxies and quasars.

We have used tunable filter observations to identify a galaxy overdensity in
the field of BR B0019-1522, with a local number density an order of magnitude
higher than that which might be expected in the field. This is among the
highest-redshift clusters detected around a radio quiet quasar. We have also
identified potential overdensities in the fields of and MRC B1256-243 and MRC
B2158-208, although deeper observations are required to confirm these
detections.

The current observations were made with the Taurus Tunable Filter, an
instrument which has now been decommissioned, on the 4 metre class
Anglo-Australian Telescope. These observations have clearly demonstrated the
success of the tunable imaging technique. The prospects for further progress
in this area are strong, as the next generation of tunable filter instruments
are now available or becoming available on telescopes such as the GTC 10-m
\citep[OSIRIS;][]{Cepa00}, SOAR 4-m \citep[BTFI;][]{Taylor10}, SALT 11-m
\citep[PFIS;][]{Smith06}, NTT 3.5-m \citep[3D-NTT;][]{Marcelin08} and the
Magellan 6.5-m \citep[MMTF;][]{Veilleux10}.

With existing telescopes, it is very difficult to extract more information
than a few emission lines and broadband photometry for the host galaxies in
these high-redshift environments. More detailed spectral information will not
be possible until the next generation of extremely large telescopes or the
James Webb Space Telescope come on line.  But there are other uses for these
observations: in particular, \citet{Bruns11} have shown that quasar
environments may act as a surrogate for studying the radiative suppression of
galaxy formation during the epoch of reionization. Interestingly, the UV
suppression reduces the star-forming galaxy counts by a factor of 2--3 but
does not suppress them altogether.  The time is therefore ripe to further
develop this promising method of investigation in order to learn about the
occurrence of high-redshift, star forming groups and the impact on these
groups by quasar activity.

\bibliographystyle{mn2e}
\bibliography{paper}

\begin{figure*}
\begin{center}
\includegraphics{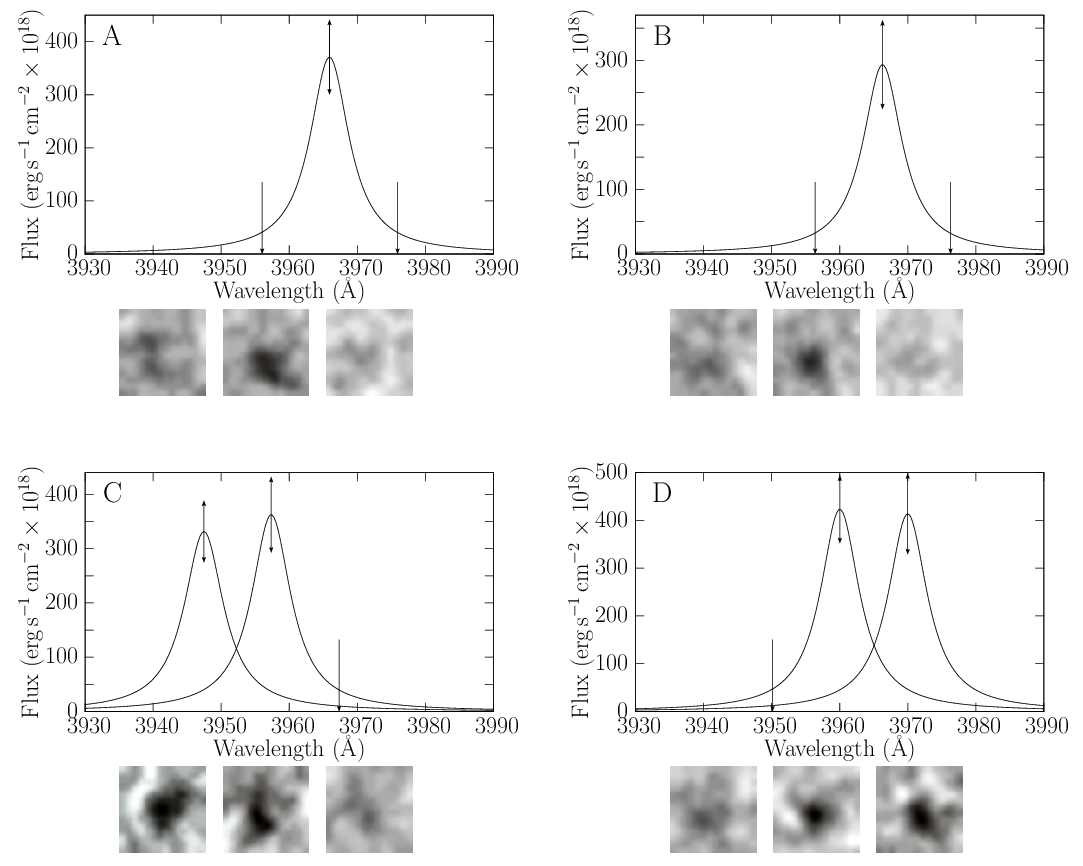}
\caption{ELG candidates in the field of MRC B1256-243. For each object, the
graph shows the flux recorded in, and the wavelength at, each etalon
transmission band. The width of the curves indicate the etalon transmission
profile. 14 pixel square thumbnail images are displayed of the objects as seen
in each band.}\label{fig:1256objects}
\end{center}
\end{figure*}

\begin{figure*}
\begin{center}
\includegraphics{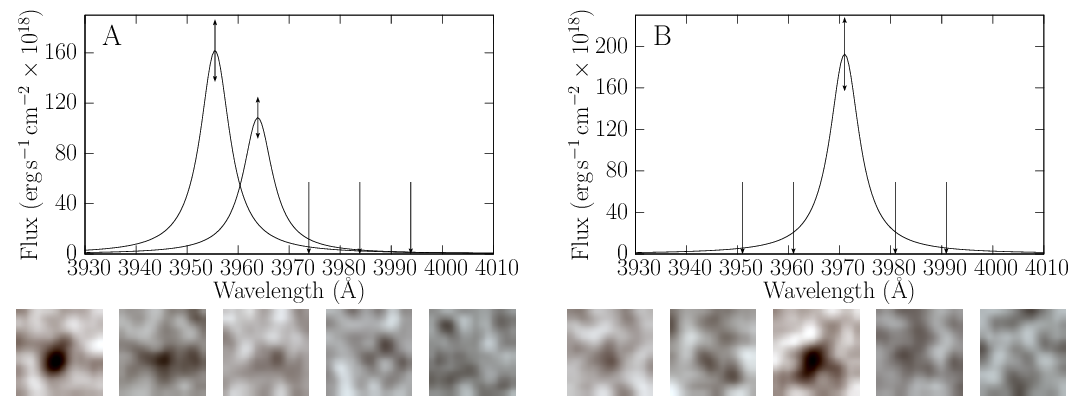}
\caption{ELG candidates in the field of MRC B2158-206. For each object, the
graph shows the flux recorded in, and the wavelength at, each etalon
transmission band. The width of the curves indicate the etalon transmission
profile. 14 pixel square thumbnail images are displayed of the objects as seen
in each band.}\label{fig:2158objects}
\end{center}
\end{figure*}

\begin{figure*}
\begin{center}
\includegraphics{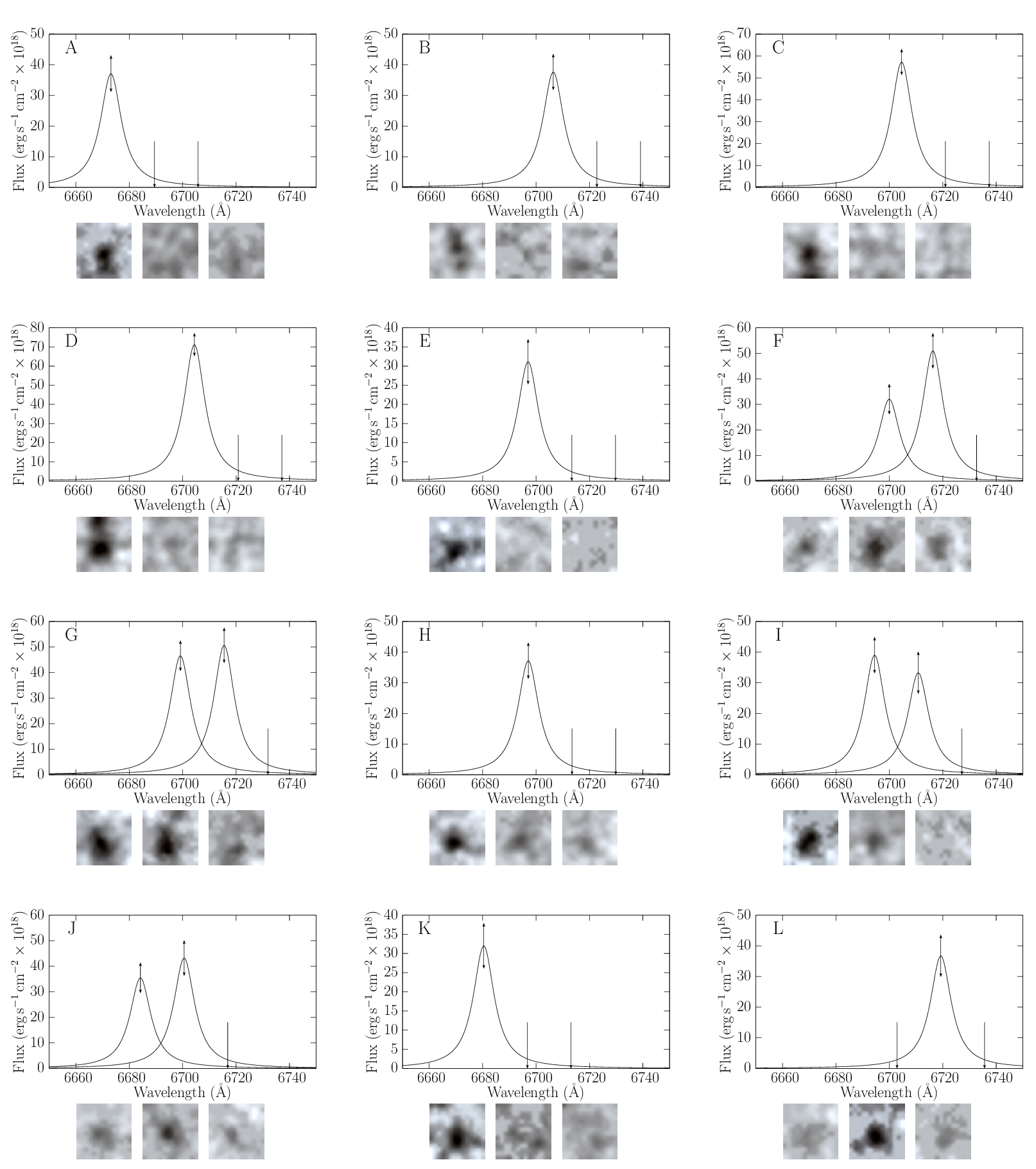}
\caption{ELG candidates in the field of BR B0019-1522. For each object, the
graph shows the flux recorded in, and the wavelength at, each etalon
transmission band. The width of the curves indicate the etalon transmission
profile. 14 pixel square thumbnail images are displayed of the objects as seen
in each band.}\label{fig:0019objects}
\end{center}
\end{figure*}

\begin{figure*}
\begin{center}
\includegraphics{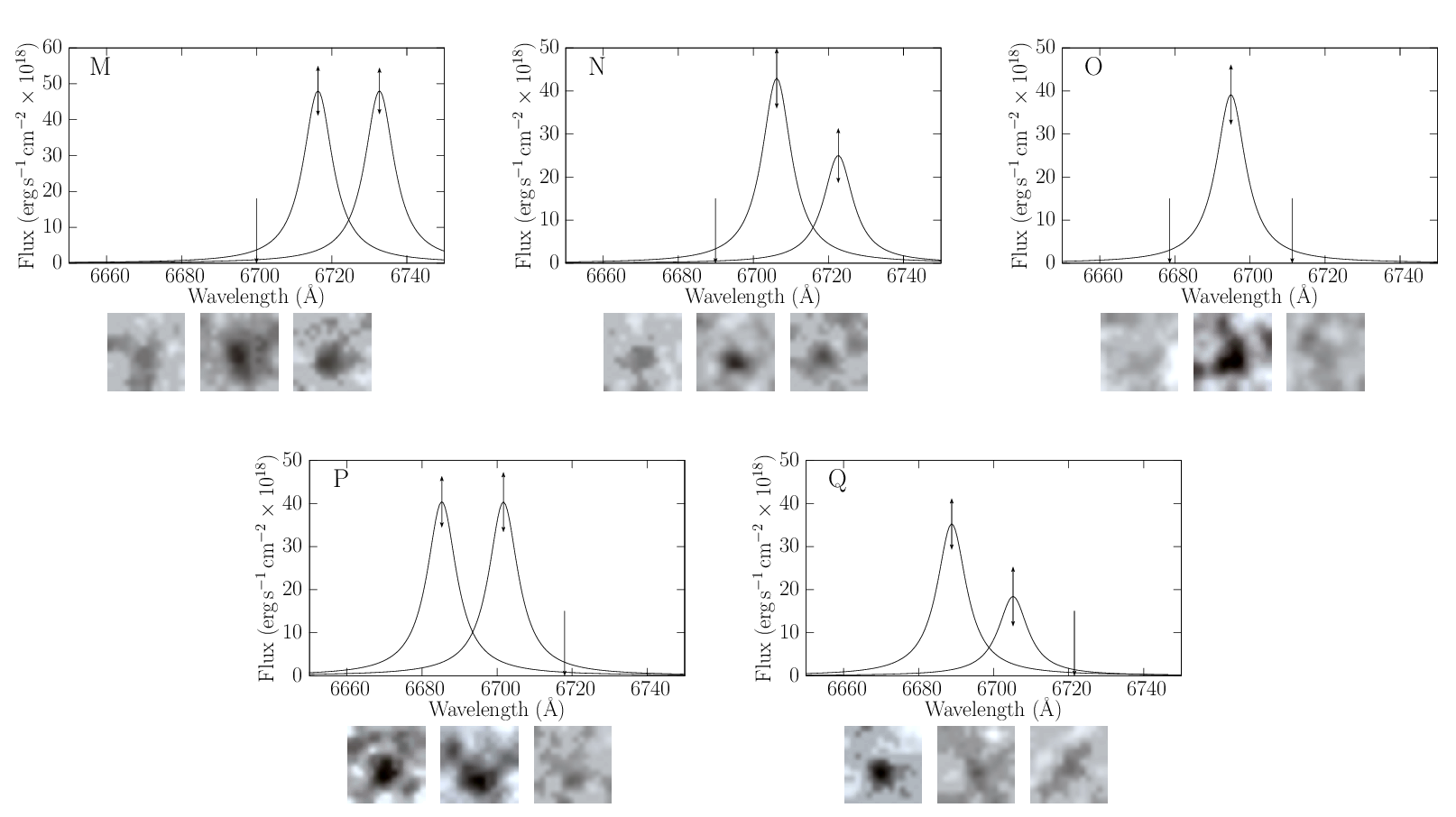}
\\
\textbf{Figure \ref{fig:0019objects} (contd).} ELG candidates in the field of
BR B0019-1522.
\end{center}
\end{figure*}

\end{document}